\documentclass[journal]{IEEEtran}
\usepackage{graphicx}
\usepackage{float}
\usepackage{caption}
\usepackage{url}

\ifCLASSINFOpdf
\else
\fi
\begin{document}

\title{A Survey of Deep Learning Techniques for Dynamic Branch Prediction}

\author{Rinu Joseph
\thanks{Rinu Joseph Department
of Electrical and Computer Engineering, College of Engineering and University of Texas at San Antonio, San Antonio Texas USA, e-mail: rinu.joseph@my.utsa.edu}
}
\maketitle

\begin{abstract}
Branch prediction is an architectural feature that speeds up the execution of branch instruction on pipeline processors and reduces the cost of branching. Recent advancements of Deep Learning (DL) in the post Moore’s Law era is accelerating areas of automated chip design, low-power computer architectures, and much more. Traditional computer architecture design and algorithms could benefit from dynamic predictors based on deep learning algorithms which learns from experience by optimizing its parameters on large number of data. In this survey paper, we focus on traditional branch prediction algorithms, analyzes its limitations, and presents a literature survey of how deep learning techniques can be applied to create dynamic branch predictors capable of predicting conditional branch instructions. Prior surveys in this field \cite{mittal2019survey} focus on dynamic branch prediction techniques based on neural network perceptrons. We plan to improve the survey based on latest research in DL and advanced Machine Learning (ML) based branch predictors.
\end{abstract}


\IEEEpeerreviewmaketitle

\section{Introduction}
\IEEEPARstart{M}{icroprocessors} with superscalar and very long instruction word (VLIW) architecture adapts instruction level parallelism (ILP) to achieve high performance. ILP executes multiple instructions of a program concurrently in one CPU cycle (clock cycle) to accelerate memory references and computation \cite{pai2000exploiting} which can result in improved performance of the processors. But one of the limitations in exploiting ILP is the high frequency of branching instructions in a program. According to Haque et al. in \cite{8880435} 15\% to 25\% of all instructions in a program are branch instructions. Three types of branch instructions are 1) jump, 2) call and 3) return. Each of these instructions can be classified as conditional and unconditional branch instructions. Branch instructions can interrupt the sequential flow of instruction execution and instead of incrementing the instruction pointer (IP) to next instruction, IP is loaded with the address of an instruction in a specified location in the memory. This interruption in execution flow can result in control hazards which causes delay by filling up the pipeline with instructions which need to be discarded subsequently.

To mitigate the issues caused by branching instructions in the code, an approach known as Branch Prediction (BP) is implemented in modern architectural designs. Branch prediction increases the instruction execution speed on processors by minimizing pipeline penalties. In BP, before the completing the execution of a branch instruction the outcome is predicted (whether the branch is taken or not-taken) and then starts executing the instructions from the predicted path prematurely. If prediction of BP process is correct then it is referred as branch hit. One of the advantage of branch hit is the effective usage of instruction pipeline. if the speculation is wrong (branch miss or branch penalty) the executed instructions from the  wrongly predicted path must be discarded and instructions from right path need to be fetched to the pipeline. This wrong prediction can causes a branch overhead which will slowdown the process \cite{kukunas2015power},\cite{calder1995system}. 

Emerging technologies like machine learning and deep learning played a vital role in the advancements in the field of computer architecture. Deep Learning,  a  branch  of  Artificial  Intelligence (AI),  uses  representation  learning  methods  to  process  the  raw  data  to  discover representations required for classification problems. DL comes with multiple levels of representations that are obtained by the composition of non-linear modules that transform representation at each level. With these composition of transformations complex functions can be learned. DL is applicable to domains like science, business and government etc \cite{lecun2015deep}. Machine learning another sub-field of AI, enables computers and machines to learn the structure of input data and enable them to make decisions based on the complex data patterns without programming them explicitly \cite{Introduc90:online},\cite{AnIntrod8:online}. The advancements in the field of ML and DL over the past few decades resulted in building more accurate systems. The increased applicability of DL models on real-world problems lead to a demand for high system performance \cite{9063049}. Researchers, system engineers and computer architects continue to design appropriate hardware which meets the computational requirement for training deep neural networks and ML algorithms on large dataset \cite{dean2018new}. While the efficiency of computer systems are improved for training purposes, there has been some research done on how to utilize ML/DL techniques to enhance the system performance. Some of the research works were focused on the area of cache replacement, CPU scheduling, branch prediction and  workload performance prediction on an x86 processor \cite{nemirovsky2018general}.

This research report on various ML/DL techniques for branch prediction is organised as follows. The following section (section 2) gives a background information about branch prediction and different branch prediction strategies and its functionalities.
In section 3 we analyze various ML/DL methodologies proposed for branch prediction. Section 5 discusses about the current limitations and scope of new advancements. This report is concluded in section 6.
\section{Branch Prediction Strategies}
To reduce the delay caused by the branch instructions in instruction fetching, instruction decoding and instruction execution, the outcome of branch i.e whether the branch will be taken or not, if it is take  what is the direction of branch is predicted before the branch is executed and decision is made. This prediction is known as Branch Prediction mechanism \cite{luo752software}. The two main techniques of branch prediction are:

\subsection{Static Branch Prediction}
It is software based prediction  strategy. This approach assumes either the branch is always take on branch is always not taken. This is a very simple and cost effective prediction technique. Since this approach does not maintains a history table of previous branch decisions it takes only less energy for instruction execution when compared to other branch prediction techniques \cite{8862216}. Following are the schemes of static (software-based) branch prediction.

\subsubsection{Single direction prediction}
In this scheme, the direction of all branch instructions will be in a similar way whether or not the branch is taken. This approach has simple implementation but low accuracy.
\subsubsection{Backward taken forward not taken (BTFT)}
The target address of backward branch is lower than the current address. This approach presumes that all the backward loops are taken and forward branches are not taken.
\subsubsection{Program based prediction} The prediction of direction of the branch instructions is based on some heuristics which are derived from opcode, operand and information of branch instructions which are executed.
\subsubsection{Profile-based branch prediction}In this scheme of static branch prediction the information from program's previous execution is used to determine the direction of branch instruction. 

\subsection{Dynamic Branch Prediction}
This hardware-based prediction strategy is based on the history of branch executions. Dynamic branch predictions are complex compared to static prediction approach but has high accuracy rate. The performance is based on the prediction accuracy and penalty rate. Different dynamic (hardware-based) branch prediction strategies are:
\subsubsection{One-bit branch prediction buffer}
In this prediction mechanism, if one assumption goes wrong branch predictor changes its prediction. i.e if a branch instruction is predicted as taken but actually the instruction is not taken, then next time predictor hardware assumes the instruction to be no taken.
\subsubsection{Two-bit branch prediction buffer}
This prediction method is similar to one-bit branch prediction buffer. The difference is that assumption is changed only after two consecutive miss-predictions \cite{Correlat72:online}. 
\subsubsection{Correlation-based branch predictor}
It is also known as two-level branch predictor. Accuracy of one-bit and two-bit branch prediction buffers are low. In correlating branch prediction accuracy is significantly improved by considering behaviour of recently executed branches for making prediction. It uses previously fetched branch target address, a local history table and a local prediction table to make accurate predictions for branch instructions \cite{Correlat72:online}. 
\section{Deep Learning/Machine Learning for Dynamic Branch Prediction}
\subsection{Perceptron Based Dynamic Branch Prediction}
\begin{figure}[!h]
\centering
\captionsetup{justification=centering,margin=2cm}
\includegraphics[width=\columnwidth]{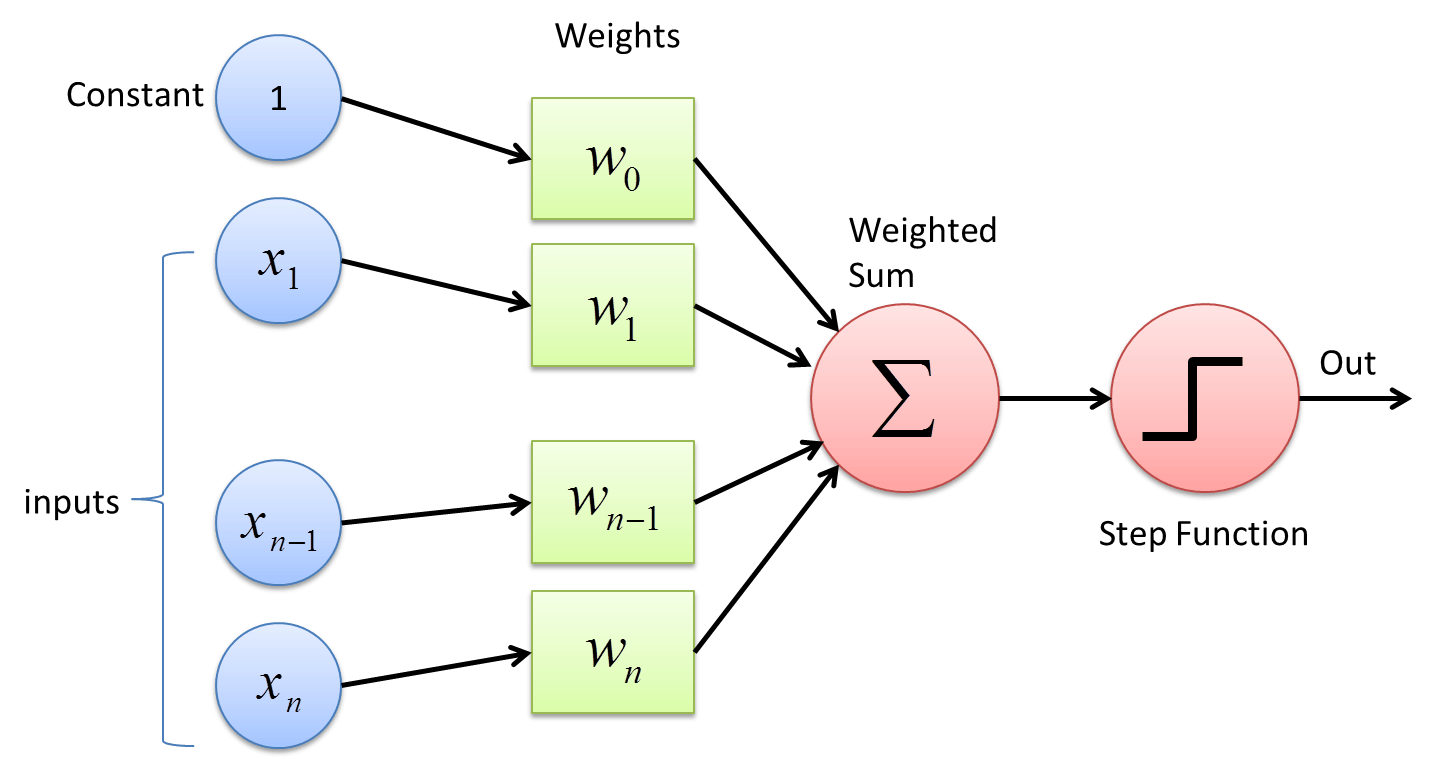}
\caption{Perceptron}
\label{fig:perceptron}
\end{figure}

In 1957 Frank Rosenblatt introduced perceptrons which was inspired by learning abilities of a biological neuron \cite{rosenblatt1958perceptron}. Perceptron is a single layer neural network which consists of a processor that takes multiple inputs with weights and biases to generate single output. As illustrated in Figure  \ref{fig:perceptron} Perceptrons work by multiplying inputs with weights and weighted sum of all multiplied values are taken and applied to an activation function.

Jimenez et al. \cite{jimenez2001dynamic} proposed a perceptron based branch prediction method which can be used as an alternative to traditional two-bit branch prediction buffers. The proposed perceptron predictor is the first dynamic predictor to use neural networks for branch prediction.The predictor make use of long branch history which is made possible by the hardware resources which can scale the history length linearly to improve the accuracy of prediction method. Basically the design space of two-level branch predictor (correlation-based branch predictor) is explored  and instead of pattern history table, table with perceptrons are used.

\begin{figure}[!h]
\centering
\captionsetup{justification=centering}
\includegraphics[width=0.8\columnwidth]{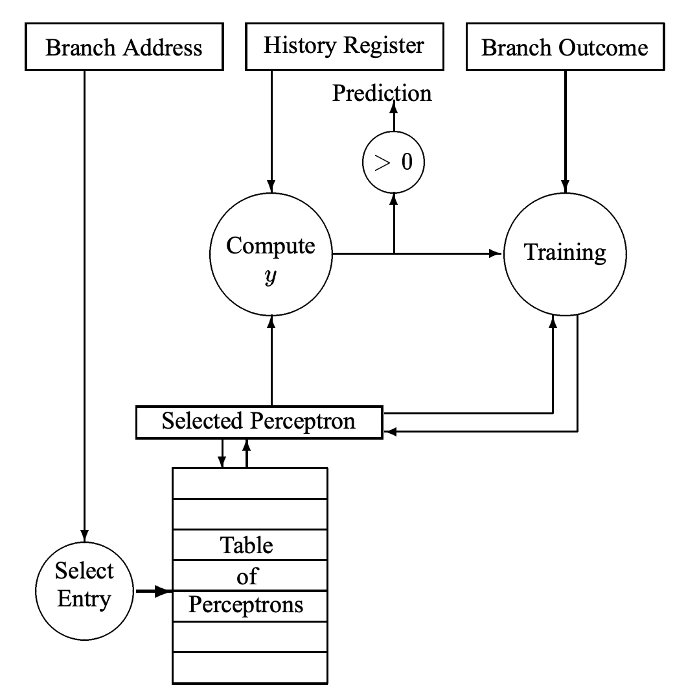}
\caption{Perceptron Predictor Block Diagram  \cite{jimenez2001dynamic}}
\label{fig:perceptron-block}
\end{figure}

In this study the branch behaviour is classified as linearly separable or linearly inseparable. Perceptron predictor's performance was better with linearly separable branches. Decisions made by perceptrons are easy to understand when compared to complex neural networks. Another factor is choosing perceptron as predictor is because of efficient hardware implementation. Other popular neural architectures like ADALINE and Hebb were used in this study but performance was low due to low hardware efficiency and low accuracy.

Figure \ref{fig:perceptron-block} is the block diagram for proposed perceptron predictor. The system processor maintains a table of perceptrons in SRAM like two-bit counters. Based on  number of weights the and budget for hardware the number of perceptrons in the table is fixed. When a branch instruction is fetched an index is produced in the perceptrons table by hashing the branch address and the perceptron at this index is moved to a vector register of weights (In this experiment signed weights are used).The dot product of weight and global history register is computed to produce output. If the output value is negative then the branch is predicted to be not taken or if the value is positive then branch is taken. When actual outcome is known, the weights are updated based on the actual result and predicted value by the training algorithm.

To achieve best performance three parameters are considered and tuned, 1) History Length, 2) Weight Representations, 3) Threshold. The increase in history length yielded more accuracy in predictions. history lengths from 12 to 62 reported best outcome for this training algorithms. As mentioned earlier the weights are signed integers and they simplify the architecture design.   

\begin{figure}[!b]
\centering
\captionsetup{justification=centering}
\includegraphics[width=\columnwidth]{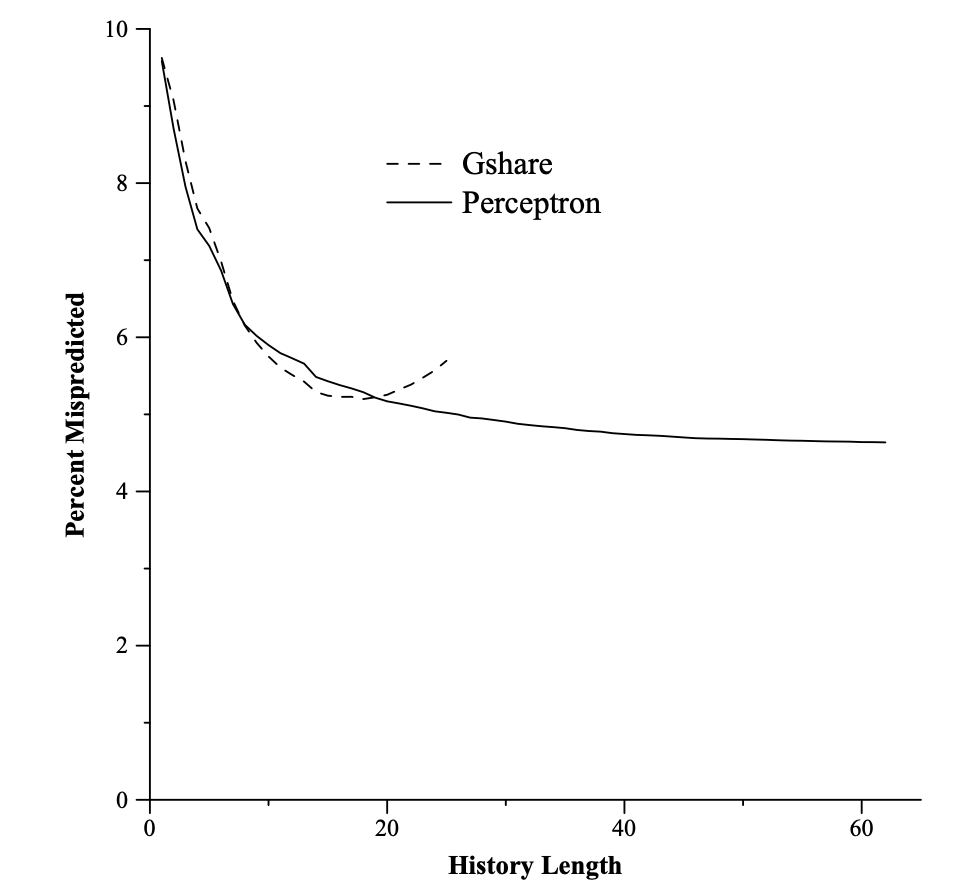}
\caption{Performance vs History Length \cite{jimenez2001dynamic}}
\label{fig:perceptron-performance}
\end{figure}

\begin{figure}[H]
\centering
\captionsetup{justification=centering}
\includegraphics[width=7cm]{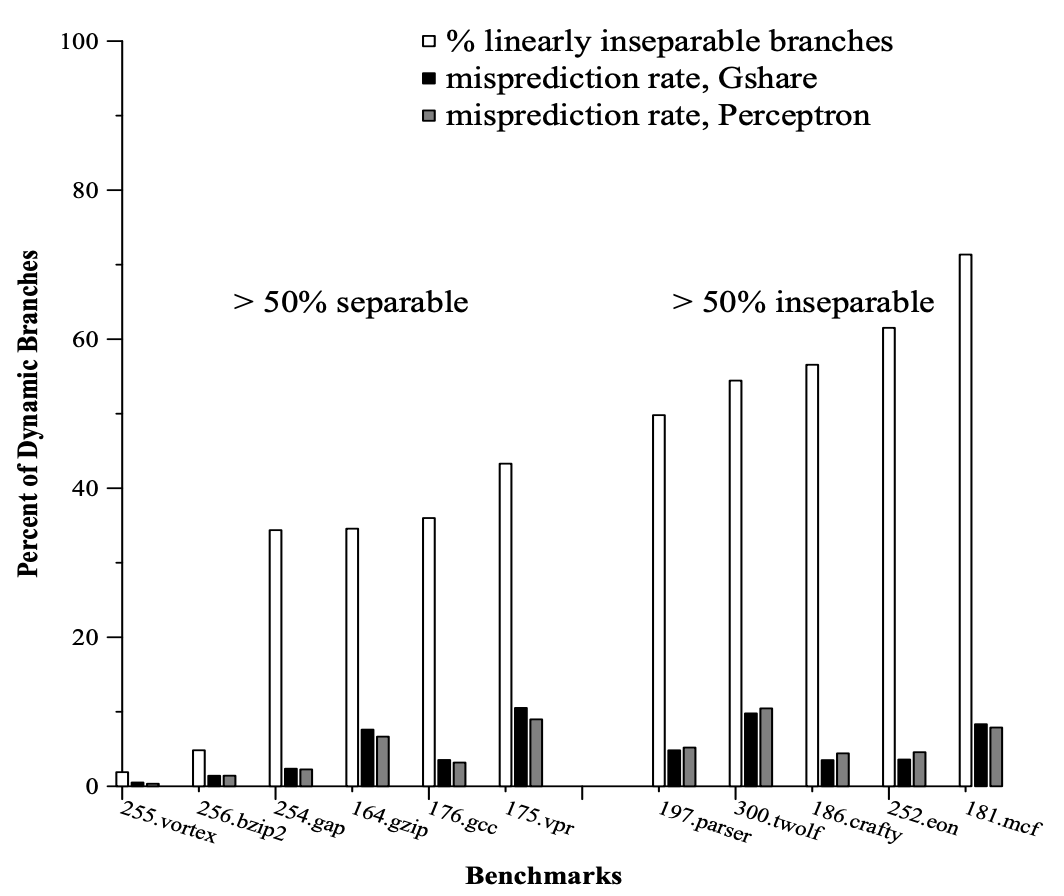}
\caption{Performance vs Linear Separability  \cite{jimenez2001dynamic}}
\label{fig:perceptron-linear}
\end{figure}

In \cite{jimenez2001dynamic} two dynamic predictors gshare and bi-mode were chosen to compare with perceptron predictor. As seen in the figure \ref{fig:perceptron-performance}, when longer histories are considered gshare shows poor performance  and perceptron predictor improves the performance. This paper is concluded by stating that although perceptrons are not capable of learning linearly inseparable behaviour of branches (Figure \ref{fig:perceptron-linear}) and they are complex when compared to two-bit counters, perceptrons can use long history tables without any exponential resources and also achieve a low misprediction rates and higher accuracy when compared to existing dynamic predictors.

Akkary et al. in \cite{1410083} proposed a perceptron based branch confidence estimator to reduce the misprediction of branch instructions. One of the factor that contribute to high performance of a system is execution of unresolved branches based on some speculations made by branch prediction techniques. Mispredicted executions negatively impact the system by taking up the resources, creating stalls in execution and also affect the power of the system because more instructions are executed when there is mis-speculation occurs. In deeper pipeline processors, pipelining gating plays a vital role in reducing wasted executions based on wrong speculative decisions made by predictors. With perceptron-based branch confidence estimator predictions provide multi-valued outputs. i.e the classify the branch instructions into “strongly low confident” and “weakly low confident”.

With confidence estimation, instruction fetching can be stalled when low confidence unresolved branch instructions are encountered. Estimation is correct when low confident branch is actually mispredicted. This correct estimation can reduce the execution wastage. When estimation is wrong (i.e low confident branch is correctly predicted) there will be a pipeline stall which can lead to performance loss.

\begin{figure}[!t]
\centering
\captionsetup{justification=centering}
\includegraphics[width=6cm]{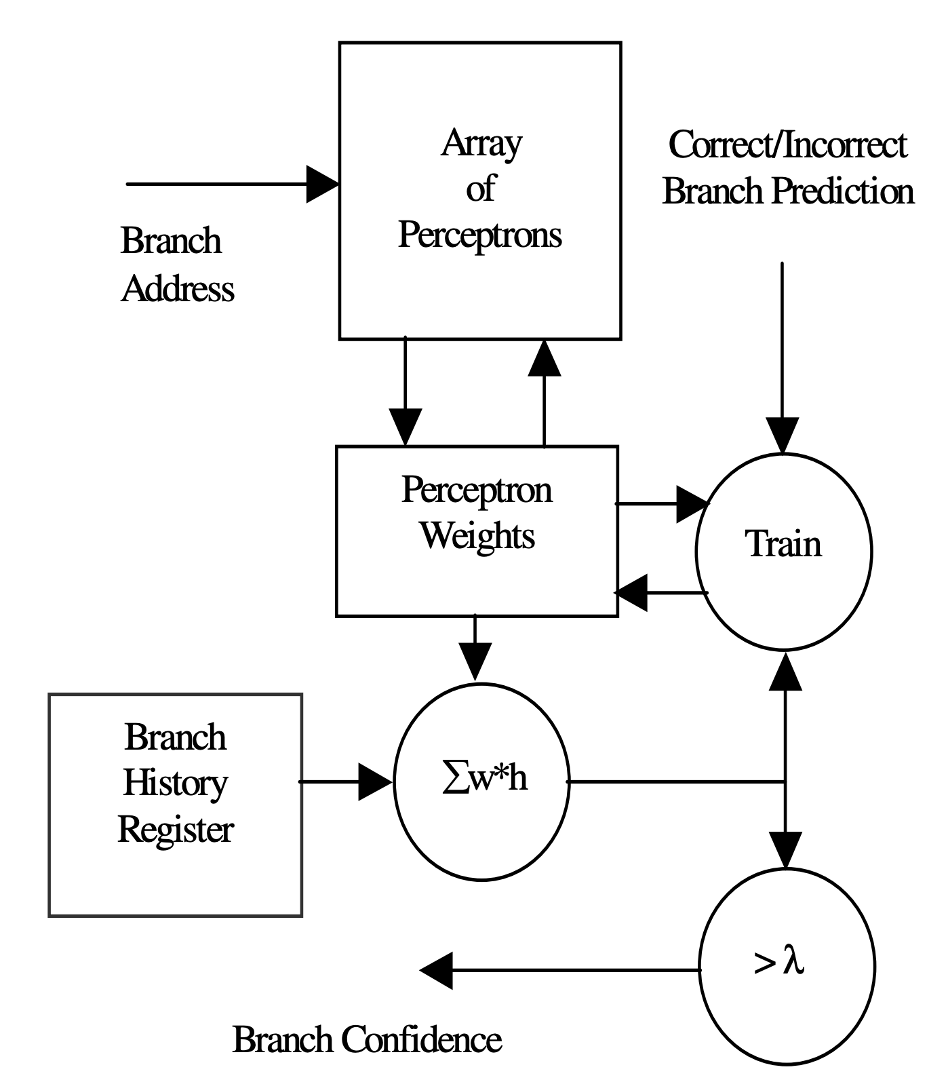}
\caption{Perceptron based confidence estimator  \cite{1410083}}
\label{fig:bce}
\end{figure}

As seen in the figure \ref{fig:bce}, index of array of perceptrons are based on conditional branch memory address. In this method global history register which contains taken branch represented as 1 and not-taken branch represented as -1 is passed as a input vector tho the perceptron. Dot product of weight vector in perceptron and this input vector is used to generate the output. As mentioned earlier the output of this approach is multi-valued. i.e, branches can be further classified as low or high confidence. The generated output is then compared against a specific threshold $\lambda$ . When output is larger then prediction can be wrong or if the output is smaller than the threshold then the prediction is more likely to be correct.

Specificity and Predictive value of a negative test (PVN) are the primary metrics used for perceptron based branch confidence estimator. Specificity is the percentage of mispredicted branches which were classified as low confidence by branch confidence estimator. PVN is the probability of correctly mispredicted low confidence branches (as a measure of accuracy).


\subsection{Branch Prediction with Convolutional Neural Networks and Deep Belief Networks}
Tarsa et al. in \cite{tarsa2019improving} says that although branch prediction attained 99\% accuracy in prediction static branches, Hard-to -predict(H2P) branches are still a major bottleneck in the system performance. Authors suggested machine learning approach to that could work along with standard branch predictors to predict H2P. They developed helper predictors mainly convolutional neural network based helpers to improve the process of pattern matching in global history data. Unlike the TAGE and preceptron based BPs, CNN based helper predictors are deployed offline.

This paper illustrated two examples of program in which the data dependency of iteration counts creates H2Ps. Variations available in data can confuse the state-of-the-art predictors but are handled by convolutional filters. The global history (instruction pointer and direction of previous branches) are converted to vector representations. The input to CNN is generated by concatenating the vectors to from a matrix for global history. CNN calculates the dot product of data vector, weight vector,filter and bias to perform the pattern matching. Score computer by each filter against the each history matrix column are passed to filter in the linear layer (similar to perceptron layer) to match the output of layer 1 in two-layer CNN. If the output of layer 2 is > 0 then the branch is taken or it is not-taken. the first layer (layer 1) in CNN identifies which instruction pointer and direction in history table correlates with the h2P's direction and layer 2 identifies which position in history contribute to branch prediction process.

This paper concluded by stating the how two-layer CNN along with the branch predictors can reduce the positional variants in global history data which can lead to branch mispredictions. They also provided some insights like passing more data like register values to network to improve the accuracy for future development of ML-based branch predictors.

Based on the work \cite{tarsa2019improving} by Tarsa et al, Zangeneh et al. developed a convolutional neural network called as BranchNet \cite{9251928} for hard-to-predict branches in 2020. As mentioned in \cite{tarsa2019improving} traditional branch predictors are updated in run-time which makes it difficult to find the correlations in branch history. Exisiting predictors cannot find correlations from a noisy history while CNN can identify those correlations correctly. This capability of CNN improves the prediction rates but requires high computational cost and large dataset for training. BranchNet can be trained offline and this models are attached to the program so that the predictors can use this model in runtime. Authors \cite{9251928}  popose CNN architecture in two-ways. 1)Using geometric history length as input to the model, 2)Sum-pooling for the compression of global history information. The designed CNN architecture is storage efficient and same latency as TAGE-SC-L.

\begin{figure}[!b]
\centering
\captionsetup{justification=centering}
\includegraphics[width=9cm]{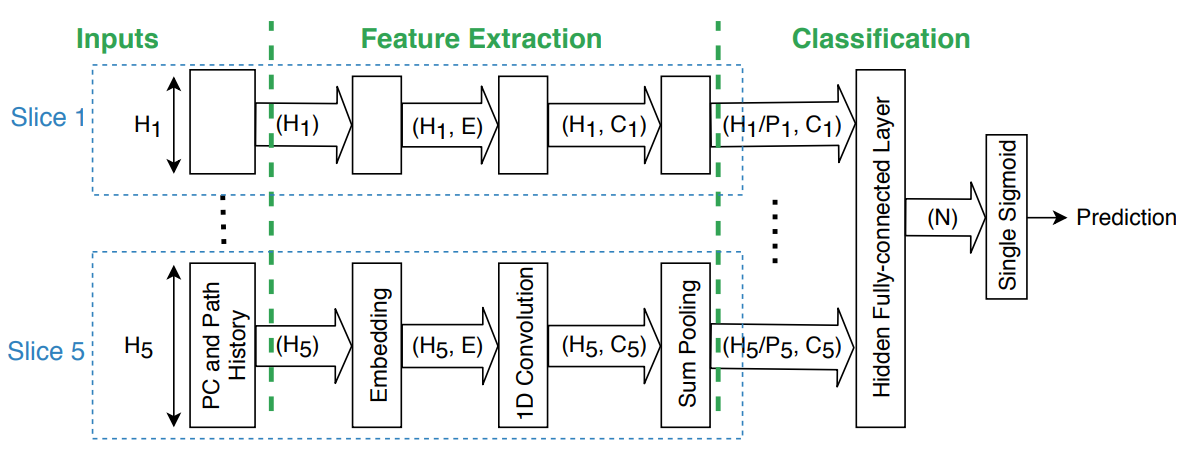}
\caption{Big-BranchNet Network Architecture  \cite{9251928}}
\label{fig:bigBrnet}
\end{figure}

\begin{figure}[!t]
\centering
\captionsetup{justification=centering}
\includegraphics[width=\columnwidth]{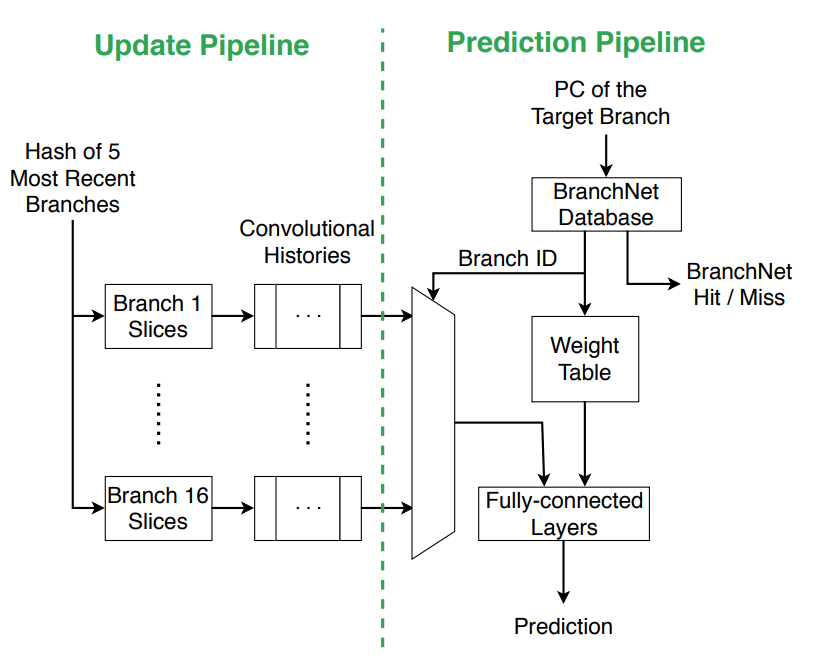}
\caption{Mini-BranchNet Network Architecture \cite{9251928}}
\label{fig:minBrnet}
\end{figure}

In this paper two-variants of BranchNet are proposed. \emph{Big-BranchNet}, consists of two fully-connected layers and five feature extraction layers (slice) (see \ref{fig:bigBrnet}). Slice consists of embedding layer, sum-pooling layer and convolutional layer for feature extraction from branch history. Each slice works on different history length to form a geometric series. Concatenated output from all slices are passed to FC layer for prediction output. Authors do not recommend the practical use of this variant as a branch predictor as this is a software model.
Another variant of BranchNet is \emph{mini-BranchNet} is co-designed with an inference engine. The architecture of min-BranchNet is similar to big-BranchNet except that it has knobs that can be tuned to reduce the latency and storage. Mini-BranchNet can work as a branch predictor.

The key achievement described in this paper is that with BranchNet deep learning network architectures can be trained in offline mode and this approach is very useful in eliminating the issues of state-of-the-art branch predictors when executed in runtime.

\begin{figure}[!t]
\centering
\captionsetup{justification=centering}
\includegraphics[width=\columnwidth]{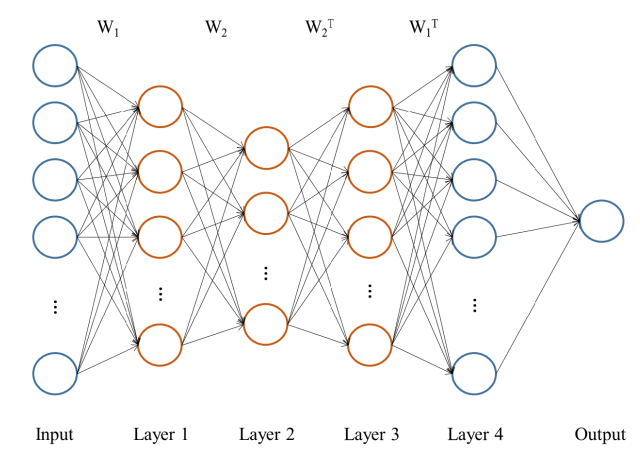}
\caption{Deep Belief Network Architecture \cite{mao2017exploring}}
\label{fig:dbn}
\end{figure}

\emph{Deep Belief Networks} (DBN) are stack of Restricted Boltzmann Machine(RBM) or Autoencoders which can be used as a solution to mitigate the vanishing gradient problem caused by backpropagation process. DBNs can be defined as a combination of probability and statistics with neural network architectures. The two layers (top) in this generative hybrid model are undirected and the layers are connected to next following layers
\cite{DeepLear85:online}. Figure \ref{fig:dbn} is commonly used DBN architecture. It has four RBM layers and an output layer. Layer 1 and layer 3 has same size. Size of input layer and layer 4 are also same. The last layer in the network architecture is a perceptron (single layer neural network). Deep Belief Networks uses unsupervised leaning techinque, where it trains the network layers to reconstruct the input data in last RBM layer. The inner layer extract the features from input to regenerate it with minimum error.

 In 2017 Mao et al. proposed the idea of using neural networks with deep belief networks for branch prediction \cite{8103746} which could outperform perceptron based BPs in terms of reducing misprediction rate. Later in 2017 \cite{mao2017exploring} Mao et al. discuss about the possibility of using advanced deep neural networks like convolutional neural network along with deep belief network for branch prediction.  Branch prediction problem is considered as a classification problem to implement different deep learning network architectures and impact of global history register (GHR), branch global address, length of program counter of the classifier on the rate of mis-speculated branch is studied . 90\% of data is used for training set and remaining 10\% is used for testing and validation purpose. For branch prediction two CNN and one DBN models were created. 

\begin{figure}[!b]
\centering
\captionsetup{justification=centering}
\includegraphics[width=7cm]{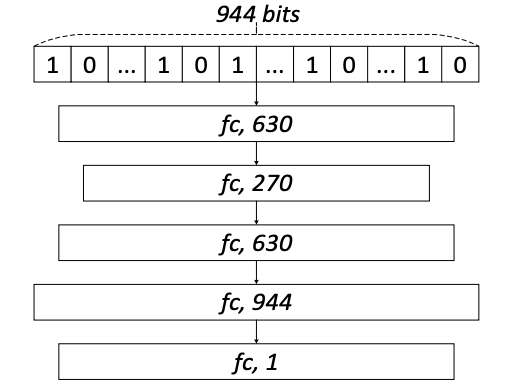}
\caption{Deep Belief Network Architecture for Branch Prediction \cite{mao2017exploring}}
\label{fig:dbn1}
\end{figure}

\begin{figure}[!b]
\centering
\captionsetup{justification=centering}
\includegraphics[width=8cm]{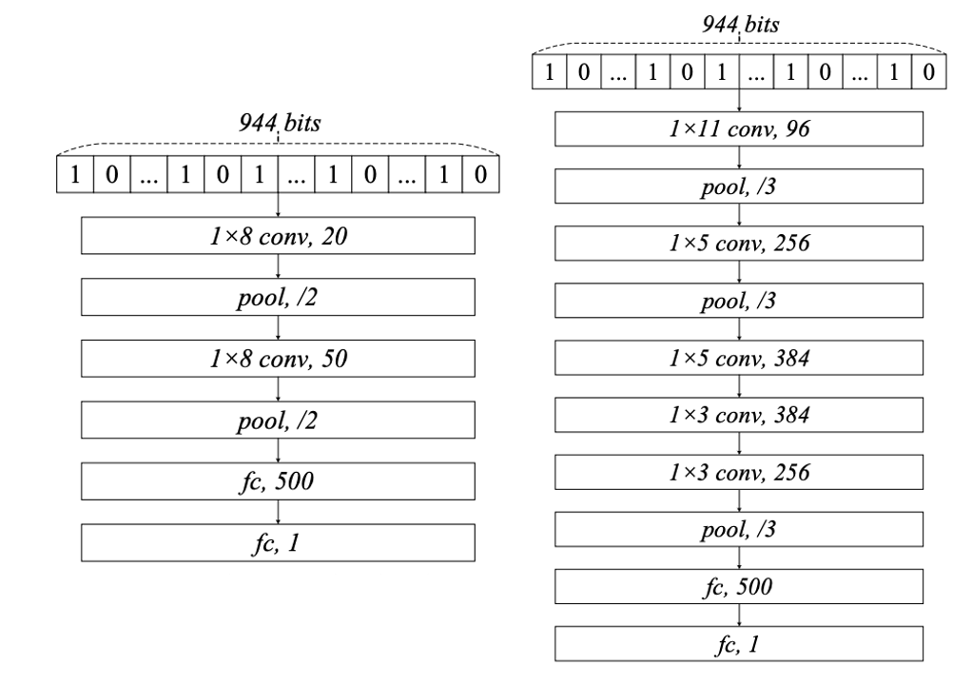}
\caption{Convolutional Network Network Architecture for Branch Prediction \cite{mao2017exploring}}
\label{fig:cnn}
\end{figure}

Figure \ref{fig:dbn1} is the DBN architecture. Neurons in the fully connected layers (FC) 1 and 3 are same (630). Layer 4 is same as the input layer. Figure \ref{fig:cnn} shows the 2 differnt CNN architectures used for this experiment. First CNN architecture is based on \emph{LeNet} and second one is based on \emph{AlexNet}. The results of these network architectures are compared against state-of-the-art branch predictors.

This paper is concluded by stating that performance of deep neural network architecture based branch predictors are better than perceptron based BPs. Deeper CNN architectures can outperform state-of-the-art branch predictors like Multi-poTAGE+SC and MTAGE+SC. When the performance of DBN and CNN are compared, CNN architecture outperformed DBN based branch predictors.

\subsection{Reinforcement Learning for Branch Prediction}
\emph{Reinforcement Learning} is a machine learning methodology in which the training algorithm estimates the error based on the penalty or reward in particular situation. The penalty is high and reward is low when the error is high. Reward is high when the calculated error is low.

In \cite{zouzias2021branch} Zouzias et al. proposed the idea of using machine learning technique Reinforcement Learning (RL) formulation in improving branch predictor designs. Today branch predictors in high-end processors are perceptron based or variants of TAGE. But there are not significant improvements in recent years to the prediction mechanism of these two variants. \cite{zouzias2021branch} suggests that branch prediction can be viewed as reinforcement learning problem because they have similar theoretical principles.We can consider branch predictor as an agent which can closely observe the control flow of the code (i.e the history of branch results) and learns a policy to improve the accuracy in future predictions. Execution environment of processor is considered as environment for branch predictor agent. Branch prediction with RL enables to explore design of branch predictor in the aspects of predictor's decision making policy, state representation and misprediction minimization strategy.   

The environment communicates with the agent (branch predictor) so that agent can choose and action from state space. Later, based on the correctness of the action selected by agent environment responds back with a reward. This reward helps to update the predictor. Normally the state space from where agent selects action contains PC address of branch, local/global history of branch instructions.

\begin{figure}[!t]
\centering
\captionsetup{justification=centering}
\includegraphics[width=\columnwidth]{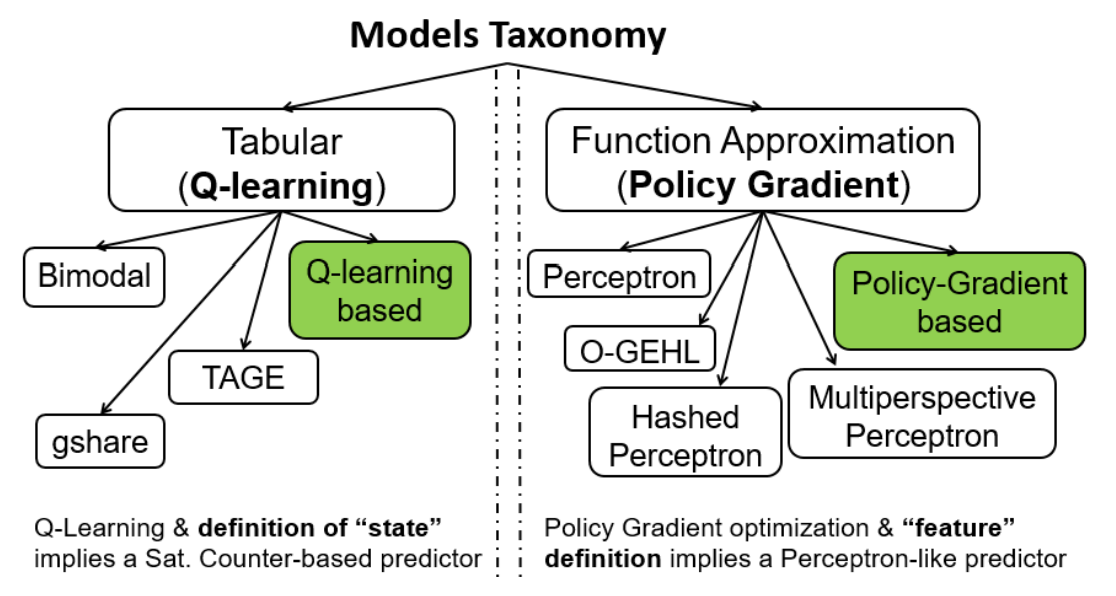}
\caption{Classification of branch predictors based on RL methods \cite{zouzias2021branch}}
\label{fig:rl}
\end{figure}

Based on the RL methods, the branch predictors are classified into two class categories 1) tabular, 2) functional. TAGE based predicors, gshare and bimodal predictors are categorized as tabular and perceptron based predictors are categorized as functional (O-GHEL, hased perceptron and multiperspective predictors). From figure \ref{fig:rl} tabular predictors can be formulated with Q-learning and functional predictors with policy-gradient methods.

This paper provide information on how to use reinforcement learning in designing future branch predictors. Key aspects to be considered for the development are:
\subsubsection{Policy} A function which can be represented as linear or non-linear model from current state to branch result. One of the advantage of this model is the easy hardware implementation . As mentioned in \cite{jimenez2001dynamic} perceptron based models cannot capture non-linear correlations. But in non-linear models micro-architectural implementations are hard and expensive.
\subsubsection{State Representation}State for branch predictions may contain branch address, local/global history and loop counters.
\subsubsection{Loss function}For branch predictors loss function (objective function) plays a vital role in achieving goal i.e reducing the mispredictions or reducing misprediction probability. 
\subsubsection{Optimization Strategy} The optimizer in RL methodology tunes learning rate to minimize the loss functions to reduce the miprediction probability. Online gradient descent optimizer is used in perceptron based branch predictors to minimize hinge loss function to adjust the learning rate.

Authors illustrated how RL methods can be applied in BP design by applying policy gradient to design perceptron based predictor. The RL based predictor is known as Policy Gradient Agent BP (PolGAg).

\section{Discussion}
Advancements in machine learning and deep learning has benefited computer architecture in many ways. Different neural networks  has enhanced the branch prediction methodolgies. In \cite{jimenez2001dynamic} and \cite{1410083} we have seen that single layer neural network called perceptron could be used for branch prediction and branch confidence estimator. Variants of state-of-the-art branch predictor TAGE and perceptron based branch predictor have latency and higher misprediction rate in hard-to-predict branches (H2P). To overcome this issue Tarsa et al. in \cite{tarsa2019improving} proposed a convolutional neural network based architecture which could be trained offline to predict H2Ps. Later in 2020 based on \cite{tarsa2019improving} work another convolutional network called BranchNet was proposed in \cite{9251928} for the same purpose. This paper designed 2 architecture Big-BranchNet and Mini-BranchNet for H2Ps.  In \cite{8103746} two architectures similar to \emph{LeNet} and \emph{AlexNet} are implemented. Although we have seen some of the researches on this topic, more study can be conducted on dynamic branch prediction and improve efficiency. Possibility of popular deep learning architectures like \emph{ResNet} and \emph{VGGnet} can be studied to implement them with branch preditors.

\section{Conclusion}
This survey focused on how different machine learning techniques and deep learning network architectures can be used to enhance the predictive ability of branch predictors. Many state-of-the-art predictors have achieved significant accuracy in branch outcome prediction. But the papers reviewed in this survey shows that, with different network architectures like perceptron, convolutional neural networks, deep, belied networks the performance of the system can be improved significantly by reducing the latency and misprediction rate.

\ifCLASSOPTIONcaptionsoff
  \newpage
\fi

\bibliographystyle{unsrt}
\bibliography{bibliography}

\end{document}